# Advanced Data Processing of THz-Time Domain Spectroscopy Data with Sinusoidally Moving Delay Lines

TIM VOGEL[*] AND CLARA J. SARACENO

*Photonics and Ultrafast Laser Science (PULS), Ruhr-University Bochum, 44801 Germany*
*Tim.Vogel-u81@ruhr-uni-bochum.de*

**Abstract:** We provide a comprehensive technical analysis of the data acquisition process with oscillating delay lines for Terahertz time domain spectroscopy. The utilization of these rapid stages, particularly in high-repetition rate systems, is known to enable an effective reduction of noise content through averaging. However, caution must be exercised to optimize the data averaging process, with the goal of significantly optimizing the dynamic range (DR) and signal-to-noise ratio (SNR). Here we discuss some pitfalls to avoid and the effect of improper data handling on the dynamic range obtainable. A free and open-source program, called *parrot* (Processing All Rapidly & Reliably Obtained THz-traces), is provided alongside this publication to overcome the discussed pitfalls and facilitate the acceleration of experimental setups and data analysis, thereby enhancing signal fidelity and reproducibility.

## 1. Introduction

In recent years, Terahertz-Time Domain Spectroscopy (THz-TDS) has gained immense popularity in research, for example due to its ability to excite and probe a large variety of physical phenomena (phonons, excitons, etc.) in condensed matter [1], [2], [3]. Additionally, THz-TDS in increasingly employed in industrial contexts, mostly for non-destructive testing such as the estimation of tablet porosity [4] or identification of rust beneath paint on steel plates [5]. Over three decades ago, a landmark publication described the use of THz-TDS to obtain phase-resolved, high-quality spectra [6]. It was promptly recognized that the acquisition speed could be significantly enhanced through the utilization of rapid, linear scanning delay lines [7], [8], and whereas several early works point to the importance of averaging and correct sampling of the delay-line position to obtain faithful data [9], [10], [11], [12], there is only limited amount of literature describing the delay between position and THz signal and its influence and their effect on the signal-to-noise ratio and dynamic range. The goal of this manuscript is to present a detailed examination of a subtle yet critical error that can arise when employing different bandwidths for the position data and the THz signal. The resulting phase delay between the two channels may result in the THz signal being associated with a shifted position data set. This phenomenon can be identified and rectified by examining the traces obtained during forward and reverse motion of the stage. In the case of a non-linear movement profile, such as a sinusoid, the previously accepted method of shifting each trace by a given amount is no longer applicable. The subsequent data processing must be conducted in a clearly defined sequence, which will also be addressed.

First, we provide a brief overview of THz generation and detection methods, as well as a selection of methods to generate the needed delay between the pump and probe arm. Subsequently, we argue that the use of rapid oscillating delay lines is ideal, in particularly considering the advent of new Ytterbium (Yb)-based, high-repetition rate lasers in a community of scientists accustomed to low repetition rate systems, thus slow step-and-settle acquisition methods. Subsequently, the source of the phase delay and a method for correcting it are described. The sample data is processed with parrot, a free and open-source software published alongside the article, which corrects for the aforementioned phase error by utilizing the information from the forward and backward traces acquired with the fast moving stage.

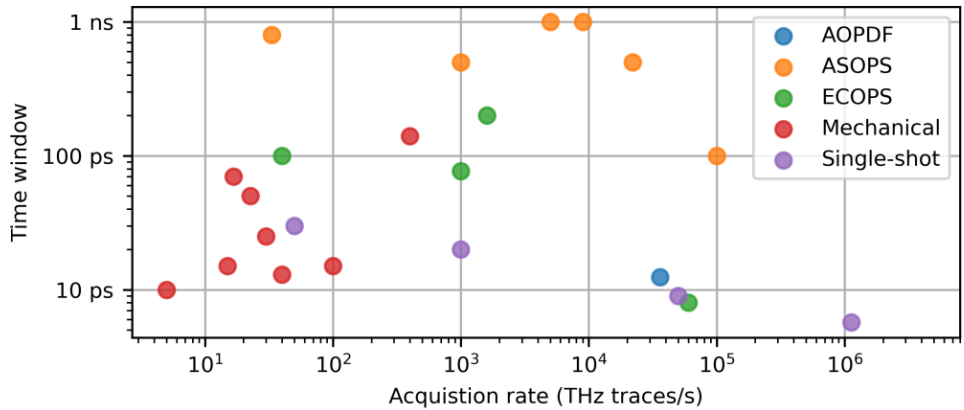

**Fig. 1** State-of-the-art results achieved with different methods to generate delay in a THz-TDS. AOPDF stands for acousto-optic programmable dispersive filter.

There are numerous methods for generating THz pulses, each with its own advantages and disadvantages. One of the most popular methods is the use of non-linear crystals [13], [14] or photoconductive emitters [15], [16]. One of the main advantages of THz-TDS compared to other techniques in this spectral range is its coherent detection principle, allowing to reconstruct both amplitude and phase of the electric field and thus extract with high sensitivity the full dielectric function of samples under study, and a variety of methods have been demonstrated to reconstruct the THz electric field in the time domain. The most commonly used methods are electro-optic sampling (EOS) [17] and photoconductive receivers [18] which can be categorized as pump-probe methods, also known as equivalent time sampling (ETS). These techniques all make use of a temporally delayed probe, most commonly by a mechanical delay line, that acts as a sampling pulse for the THz field. A pre-assumption for a highly faithful reconstruction is thus that each THz pulse is identical, i.e. that the THz pulse train is phase-stable and generally has low noise. A natural consequence of this detection method is that the acquisition rate of the THz traces is lower than the native repetition rate of the THz pulse train, thus limiting the overall acquisition speed. Further averaging over many traces may also be required to enhance signal-to-noise ratio or dynamic range, then resulting in even longer total measurement times. The resulting long acquisition times in THz-TDS remains one of the main bottlenecks to widespread implementation: for example in imaging applications for non-destructive testing, THz-TDS could achieve much higher resolution than radar methods; however real-time imaging remains out of reach for these broadband sources.

Fig. 1 gives an insight to the performance and limitation of mechanical delay lines in terms of time window and acquisition rates. Due to the inertia of the moving stage, alternative approaches allow to capture the full THz electric field at faster rates. Asynchronous optical sampling (ASOPS) [19], [20] and electronically controlled optical sampling (ECOPS) [21], for example, employ a second laser with a different repetition rate to the first laser to sample the trace without a mechanical delay line, resulting in acquisition rates in the tens of kHz. Nevertheless, high repetition rates are essential, and in most cases, modifying the repetition rate of the second system in a well-defined manner necessitates intricate stabilization schemes. This issue has been partially addressed by developing two lasers with slightly detuned repetition rates from the same cavity [22], [23]. Also, a technique called optical sampling by cavity tuning (OSCAT) eliminates the need for an external delay stage by tuning the cavity of the laser oscillator [24], [25]. Despite the rapid acquisition capabilities of these systems, no system has yet demonstrated a frequency domain peak dynamic range exceeding 100 dB at minute measurement times, whereas mechanical delay lines routinely exceed this value, most likely due to the jitter in sampling time that impedes the averaging of thousands of traces.

For sources with poor phase stability, single-shot techniques were developed, initially in the context of accelerator-based pulsed THz sources (that are typically not phase stable), but are recently being increasingly deployed to table-top laser-driven systems. These techniques map the temporal delay into another domain, for example, to the frequency domain [26], [27], [28], via a nonlinear interaction between a chirped probe pulse and the THz pulse. The current record in terms of acquisition speed is a MHz acquisition rate with a custom-made data acquisition system [29]. Whereas single-shot techniques theoretically offer the highest theoretical number of THz traces per second, offering a route to improving the shortcomings of time-scanning methods in terms of speed, several practical limitations still hinder their widespread implementation. One of them is the difficulties in data acquisition hardware's capacity to handle the amount of data or the repetition rate of the laser. Furthermore, most methods result in distorted THz traces and must meet a compromise between number of points and time window. Finally, the peak dynamic ranges observed using this method remain significantly lower than mechanical delay lines, even when significantly faster averaging is in principle possible.

The use of a mechanical delay lines thus still represents the most performant approach, offering several advantages: it remains a relatively cost-effective solution and provides high precision in position, enabling the acquisition of high peak dynamic ranges. The potential of exotic configurations like circular involute stages was explored to further enhance the rate of THz traces [30], [31], [32]. However, their calibration presents to be a challenge, and the averaging of large datasets for achieving high dynamic ranges was not demonstrated. Consequently, our attention was directed towards the more common linear delay lines. Among mechanical delay techniques two methods stand out: the "step-and-settle" approach and the continuous scanning method. The "step-and-settle" technique has minimal hardware requirements, enables high positioning precision and can be further enhanced by employing lock-in amplifiers or boxcar techniques. However, the acquisition time for a single THz trace, which depends on the step size, window size, and strength of filtering, can take minutes to complete, which additionally enhances the influence of environmental and laser fluctuations [33]. This technique is commonly adopted using very low repetition rate lasers, which themselves limit the acquisition speed of the pulses.

A solution for faster recording of THz traces with high fidelity is to adopt a continuous scanning, where the delay of the probe beam and the THz signal are recorded at the same time. By moving the motorized stage in a linear fashion, it is possible to capture approximately one trace per second, depending on the time window. Voice-coil stages offer rapid acceleration due to their low inertia, enabling oscillation frequencies of tens of Hertz and a sinusoidal movement profile. This approach has, thus far, rarely been explored for THz-TDS, with few exceptions [7], [8], [34], [35]. They are especially well suited in combination with increasingly available high average power THz sources, e.g., a lithium niobate single-cycle THz source with 66 mW and approximately 2 THz bandwidth at 13 MHz [36] or a two-color plasma based source with 640 mW and more than 20 THz at 500 kHz [37]. As these sources with wider ranges of parameters (larger bandwidth, stronger fields, etc.) at high repetition rates are becoming available, therefore this method becomes increasingly popular for detection of THz transients. However, in this detection method, special attention needs to be taken to the data acquisition of the traces to ensure fastest possible, faithful reconstruction of the pulses with highest SNR and DR. Often, record high-power THz sources fail to prove they can optimize DR accordingly, mostly due to these considerations.

To the best of our knowledge, the details of advanced data processing required to optimize the faithful detection of THz using fast, sinusoidally-moving oscillatory lines have so far not been thoroughly discussed in the literature and only a limited number of publications have highlighted the necessity for processing even for general THz-TDS data. Naftaly has highlighted the averaging in the time domain, which corresponds to vector-based averaging of the resulting spectra, effectively reducing the noise floor in the frequency domain [38, p. 53].

Withayachumnankul et al. have demonstrated an optimal order for acquiring reference and sample measurements in order to obtain more accurate estimates and standard deviations of the extracted parameters, which is particularly relevant in thin-film sensing [39]. Probst et al. demonstrated the significance of precise position data, particularly in regard to nonlinear movement profiles of a delay stage [40]. Moreover, Rehn et al. illustrated the cause of periodic sampling errors, how this error can propagate into the frequency domain, and how it can be compensated [41]. Weigel et al. implemented rapid referencing for a resonating sonotrode as a delay line for EOS and presented advanced processing techniques [42], in which a mismatch between forward and backward trace was first mentioned. Lavancier et al. also reported on a larger standard deviation, when a delay between position and THz signal is not accounted for [43]. Recently, Zhang et al. also correctly identified phase delay as the cause of the separation of forward and backward traces [44]. However, there has been no discussion of the negative effects of uncompensated phase delay, including not only a shift but also a warping of the THz traces and a reduction in the accuracy of the frequency axis. Furthermore, there has been no consideration of a straightforward, software-based solution in the time domain.

In this paper, we discuss in detail data processing steps to enhance DR and SNR in time-domain traces using sinusoidally-moving delay stages, and showcase an open-source software to the THz community that aims to circumvent common pitfalls when acquiring traces using this technique. We believe this will be of significant interest to the THz community as fast-moving delay lines become increasingly popular in the THz field together with the ongoing increase of repetition rate of THz sources.

## 2. General considerations in the design of a TDS with an oscillating delay-line

We start the discussion with general considerations and critical parameters when designing a TDS setup based on an oscillating delay line. Firstly, for each half-period of the oscillating delay line, which corresponds to one scan through the THz time window, enough pulses from the laser system must be available to fulfill the Nyquist-Shannon criterion. We note that in future implementations, this can be potentially enhanced further by employing sparse sampling or other techniques to circumvent the Nyquist limitation, however this is out of scope of this publication that aims to optimize the detection in the standard acquisition method.

The full sampling of the THz trace needs to fulfill the limitation of the voice coil frequency in accordance with the aforementioned requirement. In comparison to the conventional Ti:Sapphire amplifiers with a 1-kHz repetition rate, novel, commercially available laser systems based on Yb gain materials offer nearly the same pulse energy at 10 kHz or even 100 kHz, and systems with MHz repetition rates also start to approach the same regimes.

This increase in repetition rate can be fully exploited with a sinusoidal oscillating delay line, since the repetition rate is no longer a limitation. One reason for the limited use of this method so far, particularly in the context of common 1-kHz amplifier systems, is the relatively low number of pulses. To illustrate this, an ideal single-cycle THz pulse is simulated based on a Gaussian function,

$$g(t, \mu, \sigma) = \exp\left(-\frac{(t-\mu)^2}{2 \cdot \sigma^2}\right), \tag{1}$$

with $\mu = 0$ ps, $\sigma = 0.2$ ps and $t$ is the time. The function is twice numerically differentiated and sampled in a 50-ps time window, a value that is typically achievable by commercial voice coil stages. When sampling this simulated trace with a sinusoidal delay, the signal bandwidths at laboratory frequencies are depicted in Fig. 2a) for various shaking frequencies. The speed of the delay stage as well as other factors, which are later discussed, translate the ps-timescale ("light time") to the laboratory time frame. The corresponding laboratory frequency frame contains the signal bandwidth. As the shaking frequency is increased, the number of traces per second rises, enabling the dynamic range to be expanded through averaging. Another advantage of employing a rapid shaking rate is the spreading of the signal bandwidth, which circumvents the region affected by 1/f noise. Laser systems typically exhibit a higher noise floor at low frequencies due to environmental effects. However, this higher signal bandwidth must be supported by the laser repetition rate.

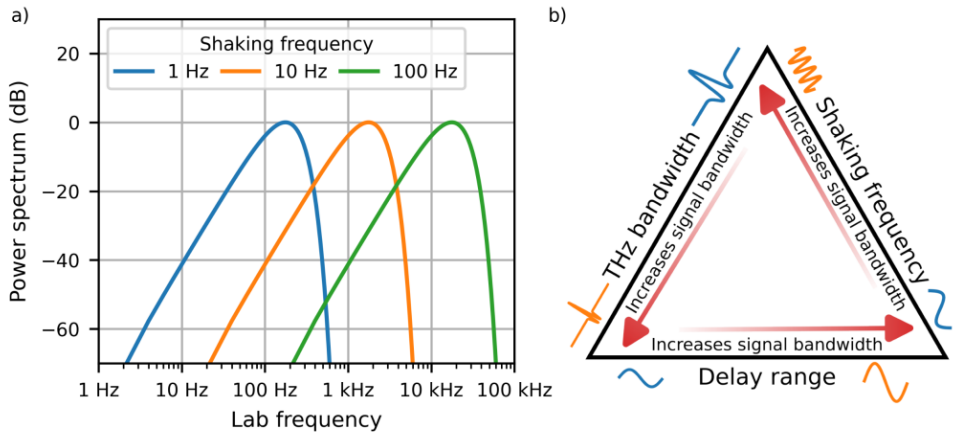

**Fig. 2**a) Signal bandwidth for simulated single-cycle THz source with a bandwidth of ~4 THz and fixed time window (50 ps) but varying shaking frequencies, resulting in larger signal bandwidths for higher shaking frequencies. b) Triangle schematic of signal bandwidth when using a sinusoidal delay line.

A lock-in amplifier, for instance, shifts the signal by the modulation frequency away from the high noise/low frequency region and subsequently demodulates the signal. However, this places a strict limitation on the integrated low-pass filter of the lock-in amplifier, which must be much lower than the modulation frequency. Typically, the low-pass filter is thus much lower than the repetition rate of the laser, which restricts the acquisition rate due to the limited signal bandwidth. It is therefore possible to include a lock-in amplifier in a continuous recording. This is particularly the case when the pump beam can be modulated rapidly enough. However, it is important to highlight that the inclusion of a lock-in amplifier is not a prerequisite for such a configuration.

An increase in shaking rate for a fixed time window and fixed terahertz signal bandwidth is not the only factor that affects signal bandwidth. The interplay between delay range, shaking frequency, and terahertz bandwidth for a given laser repetition rate can be illustrated in a signal bandwidth triangle, as shown in Fig. 2b). An increase in the delay range (for higher frequency resolution), the shaking frequency (for more traces per second) or the THz bandwidth (bigger spectroscopic coverage) results in an increased signal bandwidth, which the laser repetition rate and bandwidth of detection electronics must be capable of supporting.

Although oscillating delay lines have emerged as a highly effective method for retrieving high-quality data, the acquired data must undergo careful processing. In the next paragraphs, we present potential challenges in processing continuously recorded THz traces with a sinusoidal delay line and propose solutions to overcome them. This approach allows for the full exploitation of the data, resulting in the highest dynamic ranges and the lowest uncertainties for resonances in the frequency domain. Our newly released software to process data from continuously moving delay lines addresses these issues.

## 3. Data processing

The sinusoidal movement enables the rapid and reliable sampling of the THz trace, with the full trace observable in real time. One method for reducing the noise of this trace is to employ

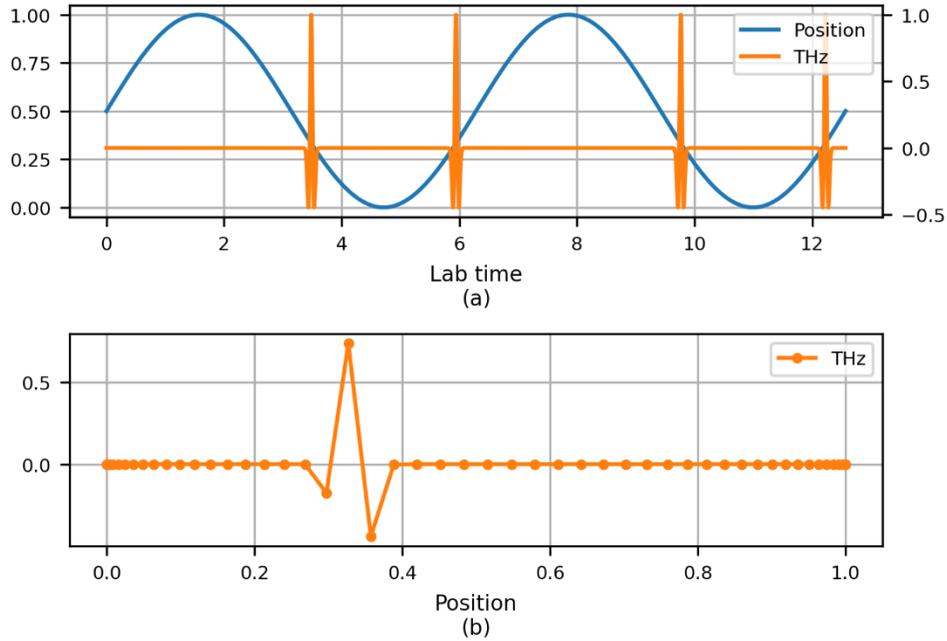

**Fig. 3** a) Sinusoidal movement of a delay stage and THz traces vs. time in the laboratory with equidistant sampling points. b) highlights, with a reduced number of sampling points in the lab time frame, the non-equidistant sampling when plotting the THz trace vs. position.

the averaging function, which is typically available on oscilloscopes. When the rising edge of the sinusoidal movement on the first channel is used as the trigger, the second channel of the THz trace remains fixed, allowing observation of a half-cycle of the delay stage and the full THz trace. As the number of averages is increased, the random noise component is reduced. This approach may be suitable for preliminary evaluation purposes, but it is important to note that there is a hidden error associated with this method, which was only observed by the authors when recording multiple THz traces in a long, continuous fashion. Moreover, the averaging of the oscilloscope prohibits the ability to perform statistical analysis on the individual traces. The following sections will discuss the continuous recording data, the associated pitfalls, and the data processing steps. The cutting, interpolating and shifting of position data versus THz data as well as systematic error correction due to additional dark measurements are implemented in the open-source software *parrot*, which we open to the community for further improvements, and can be found in [45].

### Non-equidistant spacing of continuous recorded data with sinusoidal delay

The speed of the stage is ultimately constrained by the moments of inertia of the moving component and the mirror itself. In order to maximize the number of cycles and remain below a specific maximum acceleration point, a sinusoidal movement is adapted and illustrated in Fig. 3. As the derivative of a sinusoidal modulation of position versus time also results in sinusoidal functions for velocity and acceleration, the linear stage can maximize the number of cycles while remaining below a specific maximum acceleration threshold. It has been demonstrated that a sinusoidal movement, which is non-linear, yields the highest number of cycles for a given time frame. However, this movement necessitates careful treatment of the data. A preliminary illustration of this non-linear position modulation can be observed in a simulation depicted in Fig. 3a).

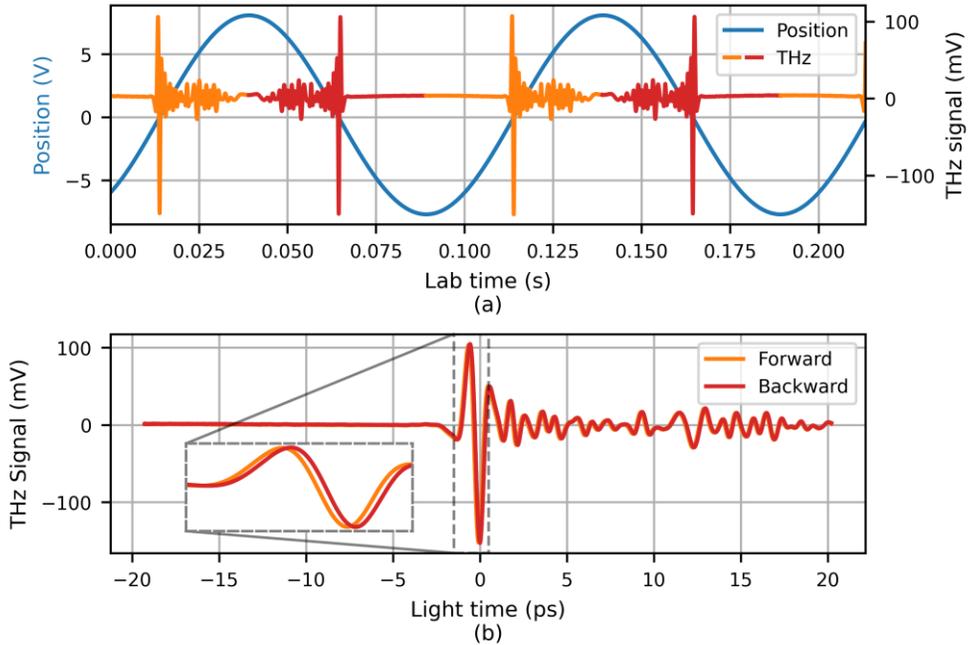

**Fig. 4**a) shows real measurement data of a THz-TDS setup with an oscillating delay line in lab time. The continuous recorded THz signal is highlighted with two different colors (orange/red) to indicate the respective movement of the stage. b) shows the THz signal vs. the recorded delay in light time. The voltage proportional to the position can be converted to light time by 50 ps/20 V for this specific model of oscillating delay line. The inset shows a zoomed-in view, displaying a shift between forward and backward traces.

The simulated data is acquired in a manner analogous to that of a data acquisition system with equidistant time-steps. When the position data is represented on the x-axis, as shown in Fig. 3b), the non-equidistant sampling of the obtained THz traces can be observed (here highlighted with a reduced number of samples). At the extremes of the position modulation, the stage moves at a slower pace, with the maximum speed occurring at the inflection point. Due to the constant sampling rate, a greater number of samples are available at the edges of the delay than in the center. The first basic consideration for applying the fast Fourier transformation (FFT) is the necessity of equidistant samples from the x-axis. This requirement can easily be met through linear interpolation.

*Compensating phase-delay between position and THz data*

Fig. 4a) depicts an actual dataset comprising multiple cycles of a delay line (APE scanDelay). The positional data is cut at both its maxima and minima. Subsequently, for each subset, the THz traces are plotted with the positional data as the x-axis values in Fig. 4b). The unexpected observation is the non-overlap of the forward (orange) and backward (red) traces. In principle, THz-TDS is a linear, time-invariant (LTI) system, and it should not matter if the oscillating delay line is moved forward or backward; and both traces would be usable, i.e. for averaging. This behavior is also not observed when the step-and-settle method is employed. If there were a non-synchronous sampling of the position channel and the THz channel, there would be a delay between these channels, which could produce an artifact. Nevertheless, modern data acquisition hardware is designed with low jitter and much higher signal bandwidth than the signals under observation, typically in the kHz range at laboratory frequencies. Another potential reason is mechanical flexure when moving back and forth. This could result in a mechanical delay when the stage already reports a future position. However, this would also

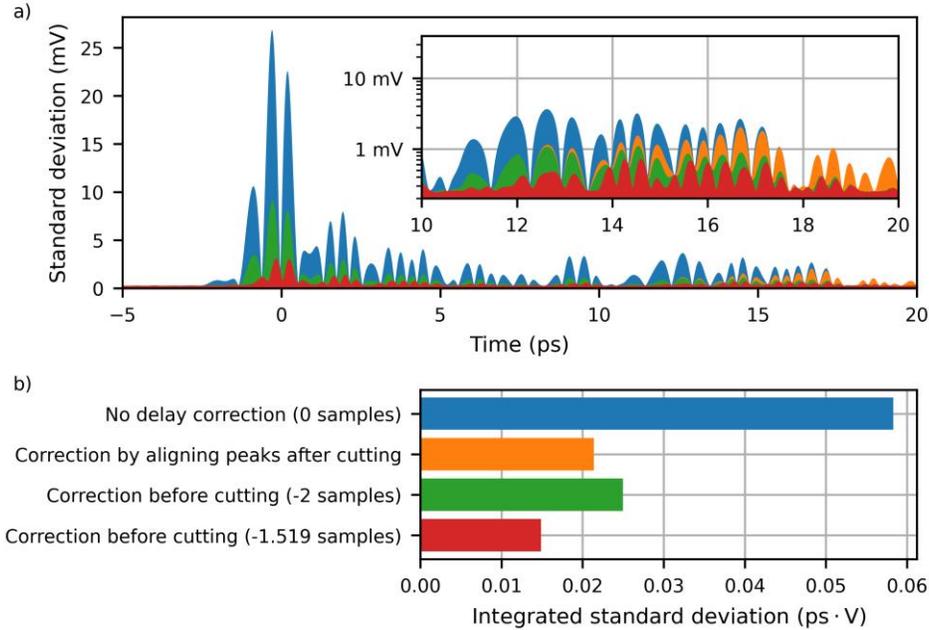

**Fig. 5** a) Standard deviation of 1202 traces captured in continuous recording mode and various correction techniques. The inset shows a zoomed-in section with a logarithmic axis. b) Integrating the standard deviation in time domain shows the benefit of correcting the delay before cutting the data, once with integer amount of time samples (green) and once with fractions of time samples (red).

depend on the speed of the stage, and we observed the same artifact for various oscillating frequencies within the same delay range.

The artifact's origin lies in the utilization of two distinct devices, namely the motion controller of the oscillating delay line and the balanced photodetector for detection inside EOS. These devices exhibit different bandwidths, resulting in a phase shift of the position versus the THz signal. The lower the bandwidth of an electronic device, the slower its reaction to signal change. At the same time, the input signals have, due to the differing bandwidths, different phase shifts and thus delays. When sampling at a sufficiently high rate and utilizing components with substantially disparate bandwidths, this phenomenon becomes more pronounced. The averaging of the data in its current form would result in the smearing out of details and a reduction in amplitude due to the misalignment of the forward and backward traces.

Fig. 5a) illustrates the standard deviation of the 1202 traces comprising the dataset presented in Fig. 4. When the delay between the forward and backward traces is ignored, the resulting standard deviation is relatively large. Several potential solutions to this problem have been identified. One potential solution is to simply ignore either the forward or backward traces and use half of the data. This presents two problematic perspectives. Firstly, a significant amount of effort has been invested in increasing the number of traces via the repetition rate. Consequently, the loss of half of them is a significant setback. Secondly, due to the differing delays between the position and THz signals, the resulting THz trace will exhibit incorrect position values and will be distorted. One potential solution is to align the peaks of each trace with each other, such that the peaks overlap. This would address the reduction in amplitude at the peak position. This approach is applicable in the case of linear scanning with a constant velocity, regardless of whether the data is aligned prior to or after cutting and using the position data as the x-axis. However, when a sinusoidal moving stage is considered, this approach becomes invalid. This can be more clearly observed in the inset of Fig. 5a). Although the

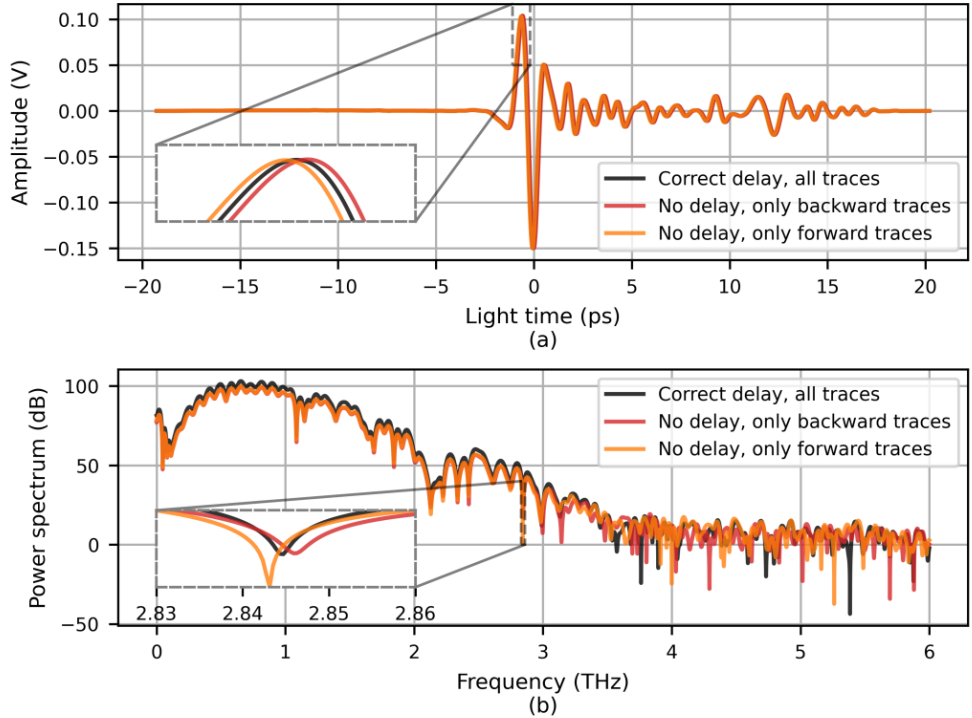

**Fig. 6** a) Averaged time domain signal, corrected with parrot before cutting the data (black) and the subset of only forward or only backward traces (red/orange) without any correction applied. The zoomed inset shows the discrepancy between the traces. b) Corresponding power spectra in frequency domain. The corrected delay has a higher dynamic range not only because it can make use of forward and backward traces. When not correcting the data, the associated warping in time domain shifts associated absorption features in frequency domain, as it can be seen in the zoomed inset.

standard deviation at the peak location is relatively low, the region from 15ps to 20ps exhibits an increase in standard deviation when the traces are shifted after the dataset is cut (orange color). This is due to the non-linear movement of the delay line which cannot be compensated by a simple shift of the x-axis after the light-time has been used for the data.

The solution to this problem is a shift of the entire positional array before it is cut and before the position data is used as new x-values. This effectively compensates for the phase delay difference between the positional data and the THz signal. A straightforward approach is to compare the difference in peak location between forward and backward traces and shift the position array accordingly. However, this would ignore the information content in the rest of the forward and backward traces, respectively. A more sophisticated approach is to utilize the integrated standard deviation over the entire time trace, with equal weights at all time samples. In the event of a delay between the position and THz signal, the full array is affected. By employing a minimization algorithm (e.g., *parrot* utilizes DIRECT [46], [47]), the integrated standard deviation can be used as an error signal, and the position array can be shifted by a new delay value, consisting of an integer amount of time samples, in order to minimize this error signal.

This already leads to an improvement in standard deviation for the entire spectrum compared to no correction, as it can be seen in Fig. 5b). While the approach is suitable for minimizing the introduced phase delay error, shifting the array by integer amount of time samples is dependent on the sampling rate of the data acquisition. In this case, the sampling rate is 50 kSa/s, which is not high enough to compensate for the phase shifts with the necessary resolution. The integrated standard deviation value demonstrates a comparable outcome to that observed when the peaks are shifted after the traces have been cut. One potential solution is to oversample the position and signal data to a greater extent than is currently the case, which would make processing the already large data files more challenging. An alternative approach, as it was developed by Denakpo et al. [48], would be to convert the data to the frequency domain and shift the phase by well-defined amounts. However, there are several considerations when employing this method. These include the computational load associated with transforming such large datasets forward and backward with the FFT for each optimization step, as well as the potential introduction of new artifacts into the dataset due to the windowing of the position data.

In contrast, a different approach is chosen in *parrot*. The original time axis of the positional data is utilized with linear interpolation. A new time axis with the same sampling rate as the original time axis is generated with a slight offset of the start time. The intermediate position data for this new time axis is requested, and subsequently, the new time axis is discarded, resulting in an effective shift of the position data by a fraction of a time sample. Fig. 5b) illustrates that this approach effectively minimizes the integrated standard deviation while compensating for the original phase delay between position data and THz signal data. This is achieved without the need for extensive computations or significant increases in memory requirements using strong oversampling of the original raw data.

Correcting this systematic error and reaching the lowest standard deviation not only allows for the highest possible dynamic range but also enables the extraction of absorption features at the correct frequency position following the application of the FFT to the data, which is often used as verification using well-defined water absorption peaks. Fig. 6a) depicts the average time domain signal following the implementation of the advanced correction algorithm in *parrot* (black), which corrects the inherent phase shift between position and THz signal by applying fractions of time samples. The other two data subsets, represented by the backward (red) and forward (orange) traces, are included for comparison purposes. These subsets also exhibit a relatively low standard deviation, but no correction is applied to the datasets. In a preliminary analysis, a shift is apparent at the peak position between the corrected dataset, as indicated by the zoomed section within Fig. 6a). The subtle warping in the forward and backward datasets is challenging to discern in the time domain. Each averaged trace is windowed (Tukey window) and zero-padded to enhance the frequency resolution. Fig. 6b) illustrates that the corrected dataset (black) exhibits a higher dynamic range than the other two subsets, due to a lower standard deviation and the inclusion of twice as many traces. Furthermore, the warping observed in the time domain is also evident in the frequency domain, as shown by the zoomed inset of Fig. 6b). This inset makes one of these shifts in the frequency domain more visible, namely the shift of a single absorption dip, which is critically important when using TDS for spectroscopy of sharp absorption features.

## 4. Potential future concepts for data processing

The methodologies utilized in this study have been validated as advantageous through the compensation of phase delay and the reduction of noise due to averaging. In future steps, sub-Nyquist approaches can be considered for further optimization. Recently, Scheffter et al. demonstrated a speed increase by under sampling the time domain traces and apply compressed sensing approaches to recover the THz electric field [49]. Furthermore, the statistical characteristics of the noise should be analyzed to identify mathematical optimization concepts to achieve the optimal balance between noise suppression and signal maximization. Initial work in that regard were already provided for example in the work from Mohtashemi et al. [50]. These and other formulations with the goal of reliable signal recovery can be accomplished by robust principal component analysis, leading to significant effort reduction in the signal acquisition. Promising results were demonstrated by Schäfer et al., showing advanced defect detection with THz-TDS systems by compensation of distortions [51]. These data processing concepts can not only be applied for an increase in dynamic range but also for super-resolution, where the resolution of a THz spectrum is no longer limited by the delay range but can be enhanced by constrained reconstruction algorithms [52].

## 5. Conclusion

In conclusion, we have summarized the use of mechanical delay lines for THz-TDS and demonstrated how they can still compete against other, even faster, sampling techniques. Their economic advantage, ease of setup, and reliable performance still serve as a good benchmark and go-to solution when setting up a THz-TDS. In particular, the fast, oscillating delay line offers an improved speed increase compared to the traditional step-and-settle technique. This makes setup alignment easier and improves the dynamic range (DR) for the same amount of measurement time.

The use of such oscillating delay lines was further motivated by the fact that even with classic amplifier Ti:Sapphire lasers operating at 1 kHz acquisition rates on the order of seconds instead of minutes can be accomplished. Moreover, the novel class of Yb-based laser system with higher repetition rates no longer impose limitations on the acquisition rate and signal bandwidths can be significantly expanded by employing faster shaking rates. This, in turn, reduces noise not only by averaging more traces within the same time frame, but also by shifting a substantial amount of the signal to higher frequencies, which are less susceptible to 1/f noise than the low-frequency region.

Nevertheless, the different bandwidths of the position signal and the detected THz signal result in subtle warping in the time domain. A post-processing method was introduced that makes use of the forward and backward traces and calculates an error signal based on the integrated standard deviation. The introduced phase shift can be compensated by minimizing the error signal by shifting the position signal with fractions of a time sample.

**Funding.** We acknowledge support by the DFG (Deutsche Forschungsgemeinschaft, German Research Foundation) Open Access Publication Funds of the Ruhr-Universität Bochum. The project "terahertz.NRW" is receiving funding from the programme "Netzwerke 2021", an initiative of the Ministry of Culture and Science of the State of Northrhine Westphalia. The sole responsibility for the content of this publication lies with the authors. Funded by DFG under Germanys Excellence Strategy – EXC-2033 – Project 390677874 - RESOLV and under Project-ID 287022738 TRR 196 for Project X01.

**Acknowledgments.** The authors thank Aydin Sezgin, Robin Löscher, and Frank Wulf for fruitful discussions.

**Disclosures.** The authors declare no conflicts of interest.

**Data availability.** Data underlying the results presented in this paper are available in [53]. The software *parrot* can be found under [45].